\documentclass[referee]{raa}
\usepackage{graphicx,times}
\usepackage{natbib}
\usepackage{amssymb,amsmath}
\bibpunct{(}{)}{;}{a}{}{,}

\usepackage{supertabular}
 \usepackage{threeparttable}
 \usepackage{longtable}
\usepackage{color}
 \usepackage{multirow}
 \usepackage{url}
 \usepackage{txfonts}
 \usepackage{algorithm}
\usepackage{algorithmic}

\usepackage[a4paper=true,pagebackref=true]{hyperref}
\hypersetup{pdftitle = The title of my PDF, pdfauthor = My name, pdfsubject= The subject, pdfkeywords = keyword1 keyword2 keyword3}
\hypersetup{colorlinks = true, linkcolor = green, anchorcolor = red, citecolor = blue, filecolor = red, pagecolor = red, urlcolor = red}
\begin{document}

  \title{A GPU based single-pulse search pipeline (GSP) with database and its application to the commensal radio astronomy FAST survey (CRAFTS)
$^*$
\footnotetext{\small $*$ Supported by the National Natural Science Foundation of China (NSFC) Programs No. 11988101, No. 11725313, No. 11690024, No.12041303 No. U1731238, No. U2031117, No.U1831131,  No.  U1831207  and  supported  by  Science  and  Technology  Foundation  of  Guizhou  ProvincePrograms No. LKS[2010]38.}
}

 \volnopage{ {\bf 2016} Vol.\ {\bf X} No. {\bf XX}, 000--000}
   \setcounter{page}{1}

   \author{Shan-ping You\iffalse（游善平）\fi\inst{1,2}, Pei Wang\iffalse（王培）\fi\inst{3}, Xu-hong Yu\iffalse（于徐红）\fi
      \inst{2}, Xiao-yao Xie\iffalse（谢晓尧）\fi\inst{1,2}, Di Li\iffalse（李菂）\fi\inst{3,4}, Zhi-jie Liu\iffalse（刘志杰）\fi\inst{2}, Zhi-chen Pan\iffalse（潘之辰）\fi\inst{3}, You-ling Yue\iffalse（岳友岭）\fi\inst{3}, Lei Qian\iffalse（钱磊）\fi\inst{3}, Bin Zhang\iffalse（张彬）\fi\inst{2}, Zong-hao Chen\iffalse（陈宗浩）\fi\inst{2}
   }

   \institute{ School of Computer Science and Technology, Guizhou University, Guiyang 550025, China ; {\it xyx@gznu.edu.cn}\\
    \and
        Key Laboratory of Information and Computing Science Guizhou Province, Guizhou Normal University, Guiyang 550001, China ; {\it
         yuxuhong@gznu.edu.cn}\\
        \and
            CAS Key Laboratory of FAST, NAOC, Chinese Academy of Sciences, Beijing 100101, China; {\it
         wangpei@nao.cas.cn}\\
            \and
           	NAOC-UKZN Computational Astrophysics Centre, University of KwaZulu-Natal, Durban 4000, South Africa
%\vs \no
 %  {\small Received XXXX XXXX XX; accepted XXXX XXXX XX}
}

\abstract{We developed a GPU based single-pulse search pipeline (GSP) with candidate-archiving database. Largely based upon the infrastructure of Open source pulsar search and analysis toolkit (PRESTO) \footnote{https://www.cv.nrao.edu/~sransom/presto/}~\citep{2001PhDT.......123R}, GSP implements GPU acceleration of the de-dispersion and integrates a candidate-archiving database. We applied GSP to the data streams from the commensal radio astronomy FAST survey (CRAFTS), which resulted in a quasi-real-time processing. The integrated candidate database facilitates synergistic usage of multiple machine-learning tools and thus improves efficient identification of radio pulsars such as rotating radio transients (RRATs) and Fast Radio Bursts (FRBs). We first tested GSP on pilot CRAFTS observations with the FAST Ultra-Wide Band (UWB) receiver. GSP detected all pulsars known from the the Parkes multibeam pulsar survey in the respective sky area covered by the FAST-UWB. GSP also discovered 13 new pulsars. We measured the computational efficiency of GSP to be $\sim$120 times faster than the original PRESTO and $\sim$60 times faster than a MPI-parallelized version of PRESTO.
 }

   \authorrunning{S. You et al. }            %author_head in even pages
   \titlerunning{Single Pulse Search Database}   % title_head in odd pages
   \maketitle

%________________________________________________ sections below
%
\section{Introduction}           %% first-level sections will be auto-capitalized
\label{sect:intro} Pulsars are rapidly rotating, highly magnetized neutron stars. Since the first pulsar was discovered in 1967, more than 2,800 pulsars
~\citep{2005yCat.7245....0M} have been detected. Pulsar detection mainly uses Fast Fourier Transform (FFT) to search in the time-frequency domain, and Fast Folding Algorithm (FFA)~\citep{1969IEEEP..57..724S} to search in the time domain. These two methods are very efficient for periodic pulsar searches. However, some pulsars have sporadic pulse trains, such as nulling pulsars and intermittent pulsars, and RRATs, which are better dealt with through single pulse detection. Fast radio burst(FRB)~\citep{2007Sci...318..777L} is a radio pulse with extremely high dispersion measure and a duration of milliseconds presumably from outside the Milky Way. FRBs can only be searched through single pulse search  at present.

With faster sampling and more simultaneous beams, the data volume of  pulsar search surveys have reached petabytes.  CRAFTS~\citep{2018Considerations}, for example, generate 6 GB/s pulsar search data streams. It is paramount to accelerate the data processing and the prioritization of the resulting pulse candidates. Due to its superior parallel processing capability and memory bandwidth, GPU has been widely used in pulsar search. The AstroAccelerate software package\footnote{https://gitlab.com/ska-telescope/astro-accelerate} with GPU implemented Fourier domain acceleration search, GPU accelerated harmonic sum of periodicity search and also using GPU implementation of single pulse detection algorithms for real-time fast radio burst searches~\citep{2020ApJS..247...56A}. Jameson and Barsdell developed a graphics processing unit (GPU) accelerated transient detection pipeline Heimdall\footnote{https://sourceforge.net/p/heimdall-astro/}.

Since millions of candidates or diagnostic plots were generated by a pulsar survey data processing, inspecting all of them by eye was time-consuming as well as a piece of work easy to made mistakes. Machine learning, particulary deep learning, have been shown to be effective in ranking the candidates~\citep{2014ApJ...781..117Z,2019SCPMA..6259507W}. For single pulses search identification, there are many tools to realise this. Karako-Argaman presented a single pulse sifting algorithm namely RRATtrap, which first grouped the detected single pulse events of nearby DMs and times, then ranked and scored the grouped candidates according to some specified rules~\citep{2015ApJ...809...67K}. Devine  presented a machine learning approach to identify and classify dispersed pulse groups (DPGs), which contain two-stages, first identified DPGs in S/N versus DMs domain by a peak identification algorithm, then used supervised machine learning for automatically DPGs classifying~\citep{2016MNRAS.459.1519D}. Michilli developed a machine learning classifier, called Single-Pulse Searcher (SPS), to discriminate astrophysical signals from a strong RFI environment ~\citep{2018ascl.soft06013M}. Pang~\citep{2018MNRAS.480.3302P} developed a two-stage single pulse search method, called Single Pulse Event Group IDentification (SPEGID), which first identified single pulse candidates as single pulse event groups (SPEGs) within DMs versus time span, then automatically classify unlabelled SPEGs as astrophysical (pulsars, RRATs or FRB) or nonastrophysical (RFI or noise) through supervised machine learning.

Pulsar search is one of the primary science  objectives for FAST. We designed a GPU-based single-pulse search pipeline with candidate-archiving database (GSP) to realize a quasi-real-time processing of the CRAFTS data. GSP implements the PRESTO de-dispersion algorithm  'prepsubband' on NVIDIA GPUs using CUDA, which performs 120 times faster than original program and 60 times faster than MPI on a NVIDIA GTX 1080 GPU. GSP incorporates a  pulsar-search database, through which  astronomers can retrieve data analysis process, and achieve centralized management of the  data processing.

 We describe GSP in section 2. In Section 3, we benchmark the GSP in test runs.
We briefly present the  processing that resulted in 13 new pulsars from CRAFTS. We present our conclusions in Section 4.

\section{Single pulse search pipeline}
\label{sect:Obs}

\subsection{Overview of the Search Pipeline}

%PulsaR Exploration and Search TOolkit(PRESTO) is a large suite of pulsar search and analysis software developed by Scott Ransom. It was primarily designed to efficiently search for binary millisecond pulsars from long observations of globular clusters (although it has since been used in several surveys with short integrations and to process a lot of X-ray data as well). PRESTO has discovered over 1000 pulsars, including approximately 400 recycled and/or binary pulsars.

We first tested GSP on pilot CRAFTS observations with the FAST-UWB receiver(270 - 1620MHz) from 2017 August to 2018 May. A series of original fits files with a fixed size were separately recorded two frequency bands of 0-1 GHz and 1-2 GHz in PSRFITS format~\citep{2004PASA...21..302H} with two polarization channels of XX and YY with 8 bit sampling, including $\sim$1800 hours of observation time (about 500TB) in 2017, and $\sim$960 hours of observation time (about 260TB) in 2018, stored in the cluster disk array and tape library of the FAST early scientific data center.

\begin{table}
\begin{center}
\caption{Frequency bands of FAST UWB drift scan data in the single pulse search pipeline}
\vspace{-5pt}
\begin{tabular}{l c c c c }
\hline
 & \multicolumn{2}{c}{\bf \text{Band1}} & \multicolumn{2}{c}{\bf \text{Band2}} \\ \hline
Frequency range (MHz)& \multicolumn{2}{c}{\text{271 to 398}} & \multicolumn{2}{c}{\text{291 to 802}} \\
Bandwidth (MHz) & \multicolumn{2}{c}{\text{128}} & \multicolumn{2}{c}{\text{512}} \\
Channels across the band & \multicolumn{2}{c}{\text{512}} & \multicolumn{2}{c}{\text{2048}} \\
Central frequency (MHz) & \multicolumn{2}{c}{\text{334}} & \multicolumn{2}{c}{\text{546}} \\
Drifting time (s) & \multicolumn{2}{c}{\text{52}} & \multicolumn{2}{c}{\text{52}} \\
\hline
\end{tabular}
\vspace{-10pt}
\label{tab:bands}
\end{center}
\end{table}

\begin{table}
\begin{center}
\caption{subband de-dispersion survey plan for Band1}
\vspace{-5pt}
\begin{tabular}{c c c c c c}
\hline
  \bf \text{No.} & \bf \text{Low DM} & \bf \text{High DM} & \bf \text{dDM} & \bf \text{DownSamp} & \bf \text{Number of DMs} \\
	      &(pc cm$^{-3}$) & (pc cm$^{-3}$) & (pc cm$^{-3}$) &	      &      \\ \hline
1 & 0 & 11.34 & 0.01 & 1 & 1134 \\
2 & 11.34 & 22.2 & 0.03 & 2 & 362 \\
3 & 22.2 & 39.05 & 0.05 & 4 & 337 \\
4 & 39.05 & 77.15 & 0.1 & 8 & 381 \\
5 & 77.15 & 153.75 & 0.2 & 16 & 383 \\
6 & 153.75 & 343.75 & 0.5 & 32 & 380 \\
7 & 343.75 & 687.75 & 1 & 64 & 344 \\
8 & 687.75 & 969.75 & 2 & 128 & 141 \\
\hline
\end{tabular}
\vspace{-10pt}
\label{tab:band1}
\end{center}
\end{table}

\begin{table}
\begin{center}
\caption{subband de-dispersion survey plan for Band2}
\vspace{-5pt}
\begin{tabular}{c c c c c c}
\hline
  \bf \text{No.} & \bf \text{Low DM} & \bf \text{High DM} & \bf \text{dDM} & \bf \text{DownSamp} & \bf \text{Number of DMs} \\
	      &(pc cm$^{-3}$) & (pc cm$^{-3}$) & (pc cm$^{-3}$) &	      &      \\ \hline
1 & 0 & 48.77 & 0.01 & 1 & 4877 \\
2 & 48.77 & 93.14 & 0.03 & 2 & 1479 \\
3 & 93.14 & 164.74 & 0.05 & 4 & 1432 \\
4 & 164.74 & 325.04 & 0.1 & 8 & 1603 \\
5 & 325.04 & 485.04 & 0.2 & 16 & 800 \\
6 & 485.04 & 592.84 & 0.2 & 16 & 539 \\
\hline
\end{tabular}
\vspace{-10pt}
\label{tab:band2}
\end{center}
\end{table}

\begin{figure*}\centering
\includegraphics[width=15cm,angle=0]{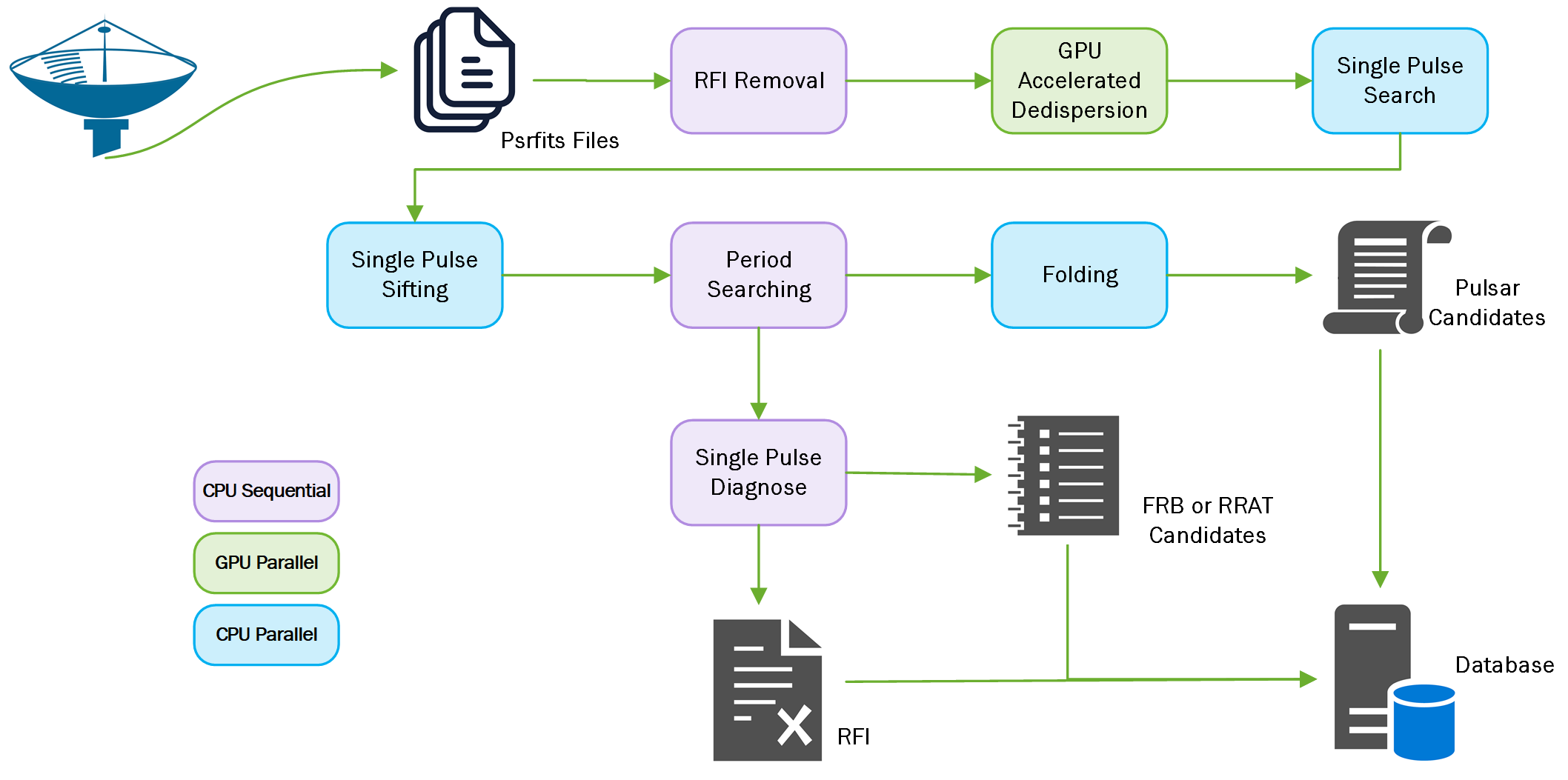}
\caption{Schematic of the CRAFTS pulsar single pulse search pipeline (GSP).}
\label{fg1:spflow}
\end{figure*}

The GSP implements GPU acceleration of the de-dispersion and integrates a candidate-archiving database. We applied GSP to the data streams from the CRAFTS survey, the data processing is performed in the FAST early scientific data center. For the single-pulse pulsar search, there are five major steps involved in processing the data: data preparing, pulsar search (RFI mitigation, de-dispersion and single-pulse search), results evaluation. See Figure~\ref{fg1:spflow} for a schematic of data processing.

In data preparation, we first merged every two of adjacent fits files for each of actual pointing, and the data stream be overlapped half of the time for the Nyquist sampling. For pulsar search, as shown in Algorithm~\ref{alg2}, under the premise that the channels and samples are determined, the number of DMs determines the calculation load of the dedispersion. In GSP, we mainly use PRESTO's 'DDplan.py' to generate the optimized DM trails. For two different frequency bands (Table~\ref{tab:bands}) , the parameters of de-dispersion trails are shown in Table~\ref{tab:band1} and Table~\ref{tab:band2}.

For the single pulse search of CRAFTS, there are two challenges to be solved in data processing. First, the de-dispersion accounts for a large proportion in the process of data processing, and secondly, it needs to identify a large number of single pulse candidates. In order to improve the search efficiency, we implemented a parallel acceleration policy to the search process, in which the longest time-consuming de-dispersion part is accelerated by GPU (see in the next section), and the single-pulse search, single pulse identification, and data folding steps were parallel executed on CPU, which saved a lot of data processing time. The entire processing flow starts with data processing. We manage with the help of a database, and save some intermediate results to the database to facilitate the later single pulse identification and diagnosis. For the single pulses candidates that failed to find its period, we used the SPS software package\footnote{https://github.com/danielemichilli/SpS} as identification tool to generate single pulse diagnostic plots to determine whether it is an astronomical signal from the universe.

\subsection{Single pulse candidate-archiving database}

\begin{figure*}\centering
\includegraphics[width=15cm,angle=0]{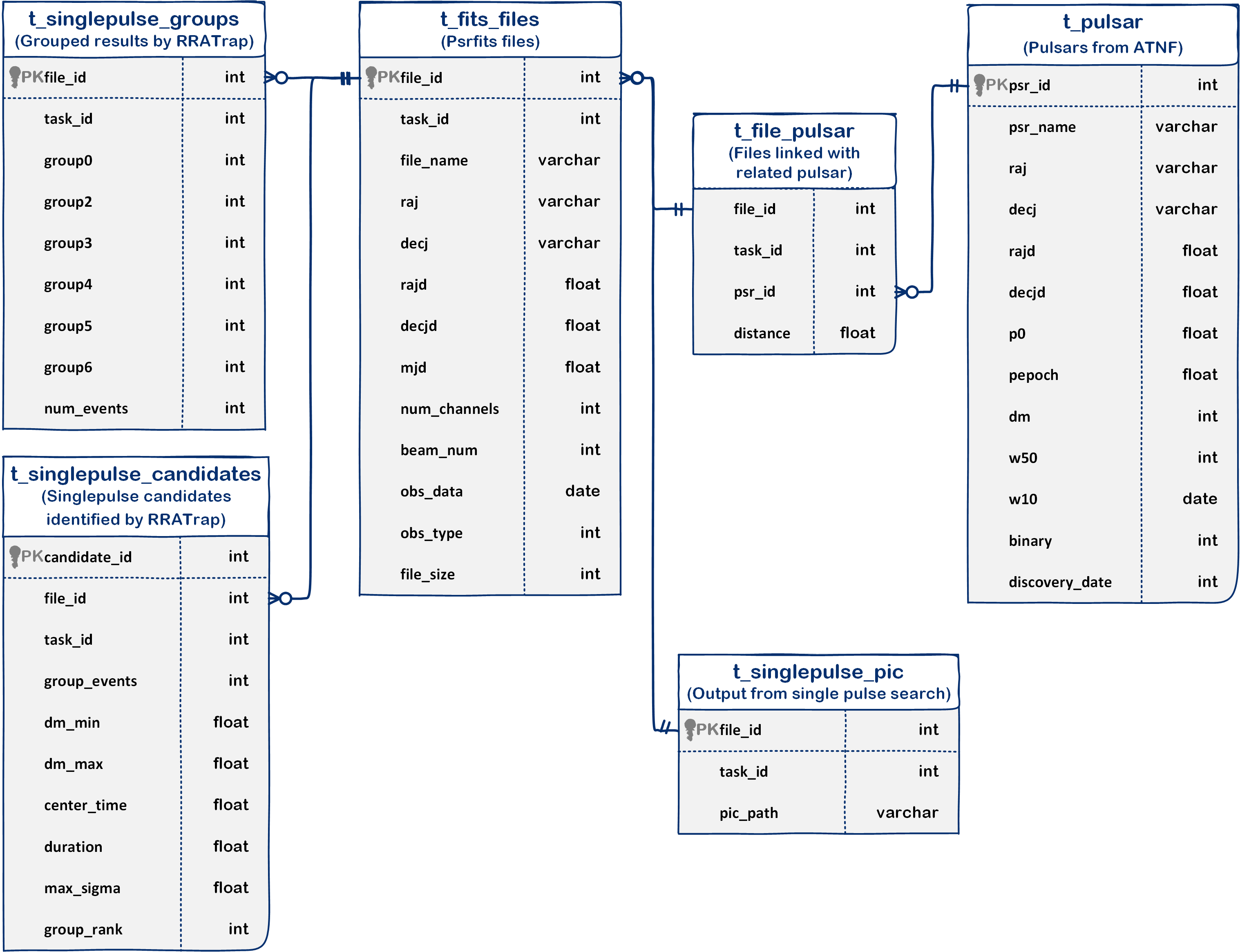}
\caption{Framework of the GSP pulsar single-pulse database.  }
\label{fg2:database}
\end{figure*}

In order to facilitate the storage, management and query of single-pulse candidates, we implemented a database based on the framework of the  mariadb~\footnote{https://mariadb.org/}. As shown in Fig~\ref{fg2:database}, the database consists of 6 tables, mainly recording the single pulse candidates identified by single pulse sifting algorithm. The database stores the psrfits file to be processed in the T\_FITS\_INFO table. For the single pulse results identified by RRATtrap, we save the overall information of each file in T\_SINGLEPULSE\_GROUPS, and store the information of each single pulse candidate, including the maximum DM , The minimum DM, the number of events of single pulse candidates, the maximum sigma, pulse appearance time, pulse duration and other information are stored in T\_SINGLEPULSE\_CANDIDATES. We saved the pulsars in the latest ATNF Pulsar Catalogue in T\_PULSAR, and compared them with DM according to the direction of the observation file, saved the results of the comparison in T\_FILE\_PULSAR, and the path of pictures generated by the pipeline was saved in T\_SINGLEPULSE\_PIC.

\subsection{GPU Accelerated de-dispertion}

For performing the blind search, we do not know the dispersion measure (DM) of pulsars, so we must make an optimized dedispersion plan and search over a wide range of trial DMs. Computing dedispersion transform is a computationally intensive task. There are three algorithms: the brute force dedispersion, the tree dedispersion algorithm~\citep{1974A&AS...15..367T}, and the Sub-band dedispersion algorithm~\citep{2012MNRAS.422..379B}. We analyzed acceleration of the parallel dedispersion algorithm on GPU, and developed a C language GPU dedispersion library. We use the sub-band dedispersion algorithm based on presto based on PRESTO's 'prepsubband'. This article conducts acceleration study and designs an acceleration program that supports multiple GPUs. %The brute force dedispersion algorithm's Algorithm~\ref{alg1} in summation over $N_{V}$ frequency channels for each of $N_{T}$ time samples and $N_{DM}$ dispersion measures gives it a computational complexity of
%\begin{equation}
%T = \mathcal{O}(N_{T}N_{V}N_{DM}).
%\end{equation}

%This algorithm was analysed previously on GPU in Barsdel(2010,2012), there are three ways in which this algorithm can be parallelized:
%\begin{itemize}
%\item with $N_{T}$ parallel threads each compute the sum of a single time sample over every channel sequentially;
%\item with $N_{V}$ parallel threads cooperate to sum each time sample in turn;
%\item with $N_{T} \times N_{V}$ parallel threads cooperate to complete the entire computation in parallel.
%\end{itemize}

%\begin{algorithm}
%\caption{Pseudocode of the dedispersion algorithm}
%\label{alg1}
%\begin{algorithmic}
%\REQUIRE data array with $N_{V} \times N_{T}$, $N_{DM}, N_{V}, N_{T}$
%\ENSURE result array with $N_{DM} \times N_{T}$
%\STATE dSample = 0
%\FOR{$dm = 0\rightarrow N_{DM} $}
%\FOR{ $chan = 0\rightarrow N_{V} $}
%\FOR{ $sample = 0\rightarrow N_{T} $}
%\STATE dSample += data[chan][sample + $\Delta$(chan, dm)]
%\ENDFOR
%\ENDFOR
%\STATE result[dm][sample] = dSample
%\ENDFOR
%\end{algorithmic}
%\end{algorithm}

In presto prepsubband, the input of dedispersion was stored in two preprocessed channelized time-series matrix with $numpts \times numchan$, where $numpts$ is the number of time samples per read, $numchan$ is the frequency channels, the delay of each channel was pre-calculated and stored in a $numdm \times numchan$ matrix, $numdm$ is the number of DMs to dedisperse, the results after dedispersion was stored in a matrix with $numdm \times numpts$, the algorithm was shown in Algorithm~\ref{alg2}.

\begin{algorithm}
\caption{Pseudocode of the dedispersion algorithm of presto prepsubband}
\label{alg2}
\begin{algorithmic}
\REQUIRE data array with $numpts \times numchan$, delays array with $numdm \times numchan, numpts, numchan$
\ENSURE result array with $numpts$
\FOR{$ii = 0\rightarrow numpts$}
\STATE result[$ii$]=0
\ENDFOR
\FOR{$ii = 0\rightarrow numchan$}
\STATE $jj=ii+delays[ii] \times numchan$
\FOR{ $kk = 0\rightarrow numpts - delays[ii] $}
\STATE result[$kk$] += lastdata[$jj$]
\ENDFOR
\STATE $jj=ii$
\FOR{ $kk = numpts - delays[ii]\rightarrow numpts$}
\STATE result [$kk$] += data[$jj$]
\ENDFOR
\ENDFOR
\end{algorithmic}
\end{algorithm}

Since presto was originally developed based on CPU serial programs, the code was gradually optimized to support parallel programs of OPENMP and MPICH, but the overall program was based on CPU. Therefore, to achieve GPU accelerated prepsubband calculation, it is necessary to redesign the de-dispersion acceleration algorithm based on the GPU programming model. From Algorithm~\ref{alg2}, the loop of time sampling is divided into two parts by the delay matrix, the part with delay less than $numpts$ minus $delays$ gets data from lastdata matrix, and the rest gets data from data matrix. This type of design will resulting in divergence on the GPU, thereby reducing the performance of dedispersion. To eliminate the divergence, we merged the two matrices and get a new matrix with $ 2 \times numpts \times numchan$. As shown in Algorithm~\ref{alg3}, we revised the dedispersion algorithm with $numpts \times numdm$ threads, each thread compute the sum of a single time sample over every channel sequentially. Unlike Algorithm~\ref{alg2}, for each thread, we add the sum of sample over channels to a variable which stored in the GPU registers, this reduced the number of accesses to global memory.

\begin{algorithm}
\caption{Pseudocode of the parallel dedispersion algorithm on GPU}
\label{alg3}
\begin{algorithmic}
\REQUIRE data array with $ 2 \times numpts \times numchan$, delays array with $numdm \times numchan, numpts, numchan$
\ENSURE result array with $numpts$
\STATE tid = blockIdx.x * blockDim.x + threadIdx.x
\IF{$tid < numpts \times numdm $}
\STATE resultvar = 0
\FOR{ $ii = 0\rightarrow numchan $}
\STATE jj=ii*2*numput+(tid mod numpts)+delays[$\lfloor tid / numpts \rfloor$ * numchan + ii]
\STATE resultvar += data[jj]
\ENDFOR
\STATE result[tid] = resultvar
\ENDIF
\end{algorithmic}
\end{algorithm}

For input data preparation, in order to be compatible with the original presto, the original code has not been changed, which can ensure that the parameters of the dedispersion input are consistent with the original parameters, thereby ensuring that the result of the GPU computation is the same as that of the original CPU.

\section{Pipeline Tests}

In this section, we present the test of GSP to the data streams from CRAFTS survey.

\subsection{Experiments of GPU Accelerated De-dispertion}

The FAST Ultra-Wide Band (UWB) receiver achieves a better than 60-K receiver temperature covering 270 MHz–1.62 GHz. The beam-passing time of a drift scan equals approximately 52s at frequency of Bands in Table~\ref{tab:bands}. With proper weighting, the equivalent integration time per beam will be around 65s. The recorded FAST data stream for pulsar observations is a time series of total power per frequency channel, stored in PSRFITS format ~\citep{2004PASA...21..302H} from a ROACH-2 \footnote{https://casper.ssl.berkeley.edu/wiki/ROACH-2\_Revision\_2} based backend, which produces 8-bit sampled data over 256 - 4k frequency channels at 98 $\mu$s cadence. For the GPU de-dispersion test in this work, we carried out an experiment by adding simulated pulses into real data. The injected pulses were generated assuming various DMs and then sampled in time and frequency in exactly the same fashion as those of the FAST data. Since the DM of a pulsar is not known a priori, we created de-dispersed time series for each pseudo-pointing over a range of DMs, from 0 to $\sim$1000 pc $cm^{-3}$,  which is a factor of three to four larger than the maximum DMs predicted by the NE2001 model ~\citep{2002astro.ph..7156C} and YMW16 model ~\citep{2017ApJ...835...29Y} in the low Galactic latitude regions of the survey. The step size between subsequent trial DMs ($\Delta {DM}$) was chosen such that over the entire band t$_{\_ \Delta {DM}}$ = t$_{\_ chan}$. This ensures that the maximum extra smearing caused by any trial DM deviating from the source DM by $\Delta{DM}$ is less than the intra-channel smearing. The set of trial values is chosen using the 'DDplan.py' tool from PRESTO and is shown in Table~\ref{tab:band1} and Table~\ref{tab:band2}. We tested a total of 48 test files, and the number of DMs Start from 30 steps and test to 900 steps. Each test process tests CPU single-thread, CPU 16-thread OMP acceleration (implemented with -ncpus 16), and 15-process mpiprepsubband (because mpiprepsubband must meet the number of processes n-1 times the number of cpu cores), and gpuprepsubband acceleration tests.

\begin{figure*}
\begin{tabular}{ll}
\includegraphics[width=8cm,angle=0]{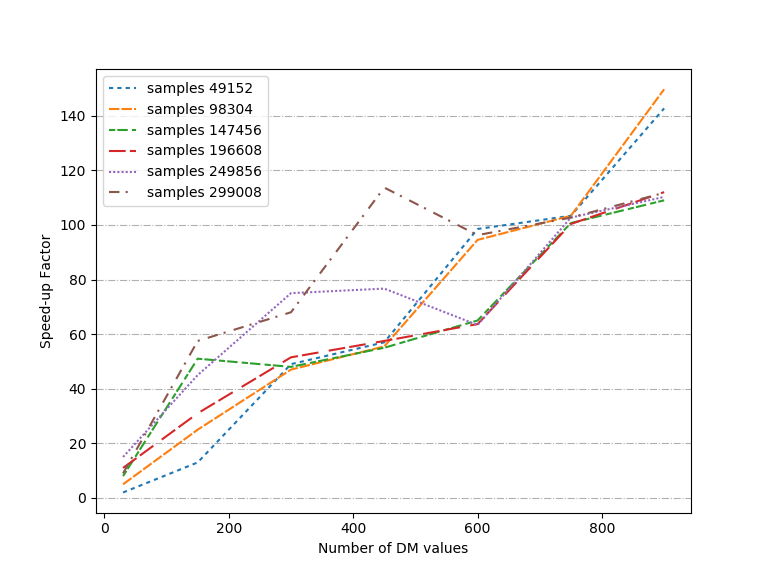} &
\includegraphics[width=8cm,angle=0]{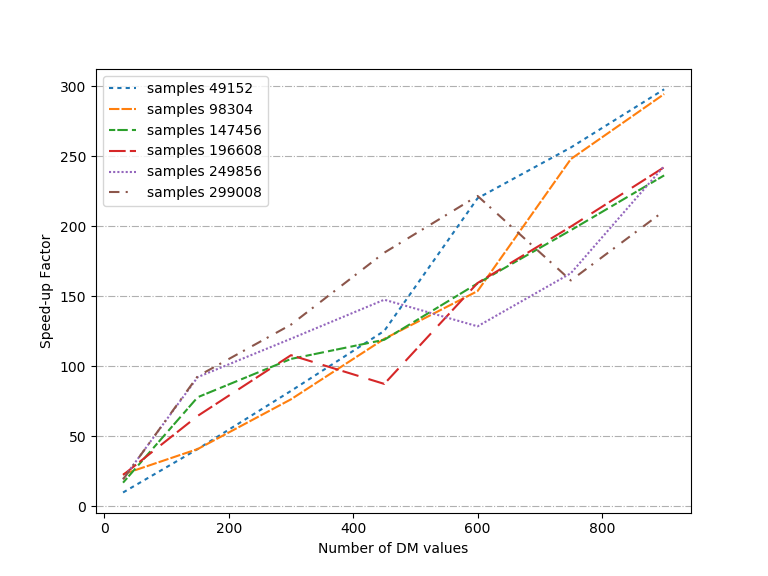} \\
\includegraphics[width=8cm,angle=0]{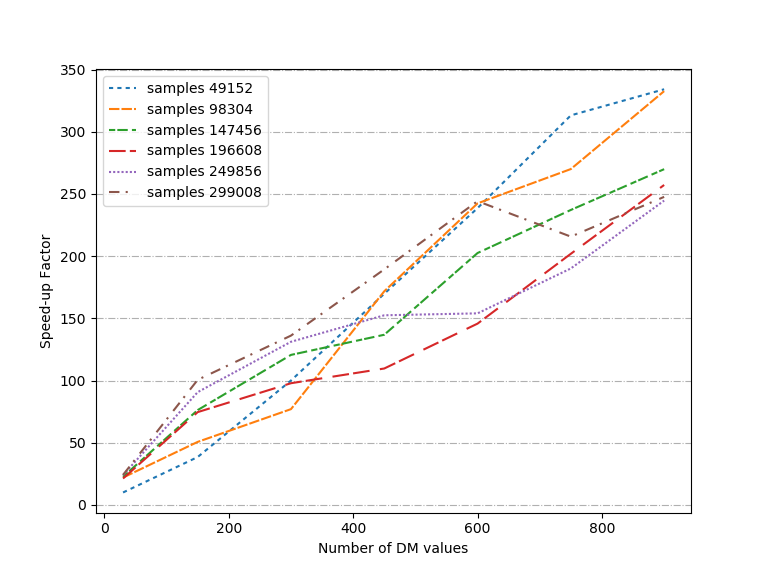} &
\includegraphics[width=8cm,angle=0]{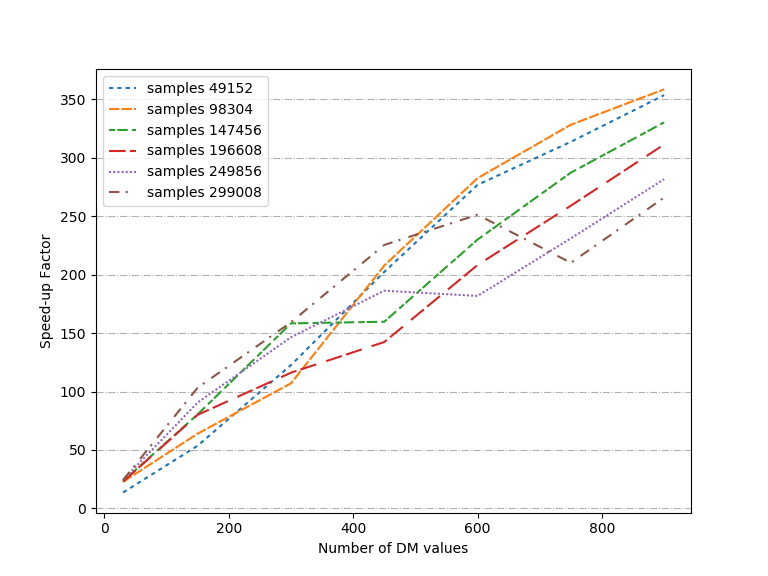} \\
\end{tabular}
\caption{When the number of channels is 256, 512, 1024, and 2048, the gpuprepsubband is compared with the acceleration effect of prepsubband running 1
processes. It can be seen from the figure that as the number of tested DMs increases, the acceleration effect becomes better and better; when the number of
DMs is relatively large , The acceleration effect of short files is better than that of long files.}
\label{fg2:cpugpu}
\end{figure*}

\begin{figure*}
\begin{tabular}{ll}
\includegraphics[width=8cm,angle=0]{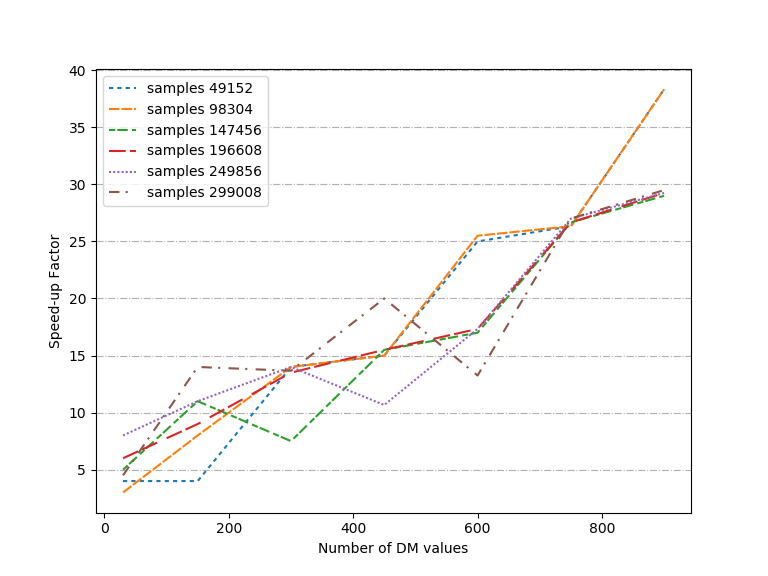} &
\includegraphics[width=8cm,angle=0]{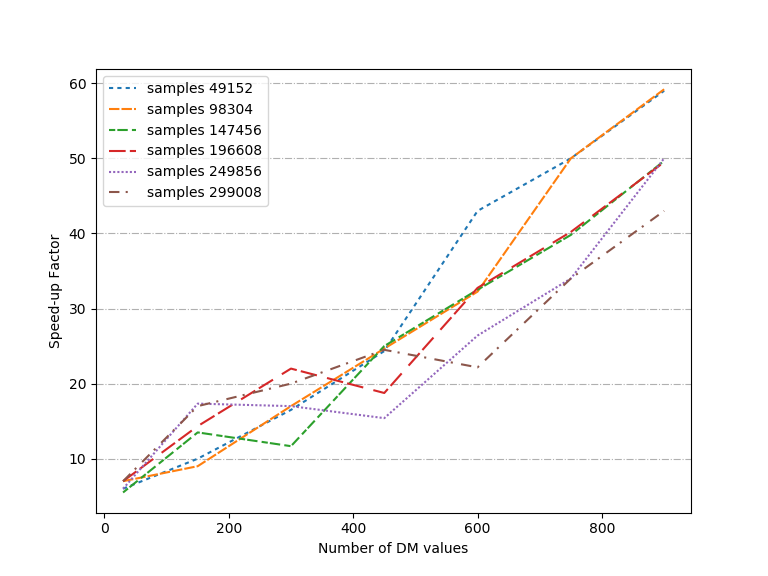} \\
\includegraphics[width=8cm,angle=0]{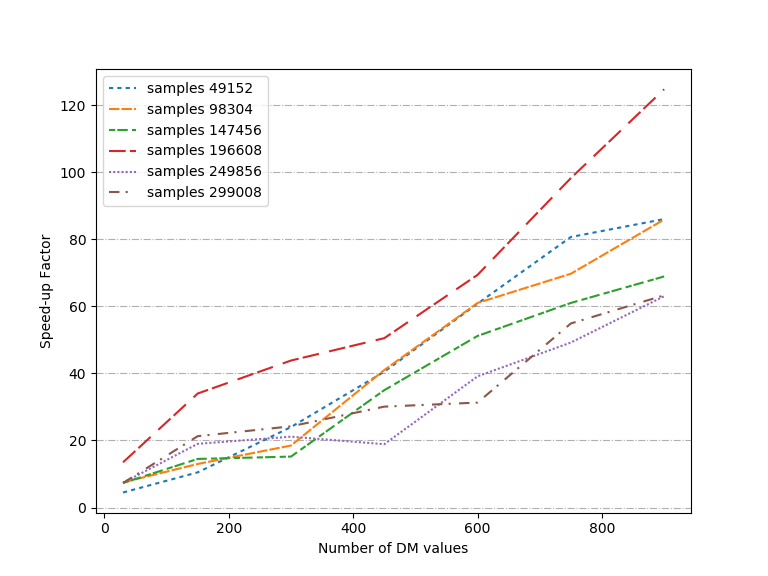} &
\includegraphics[width=8cm,angle=0]{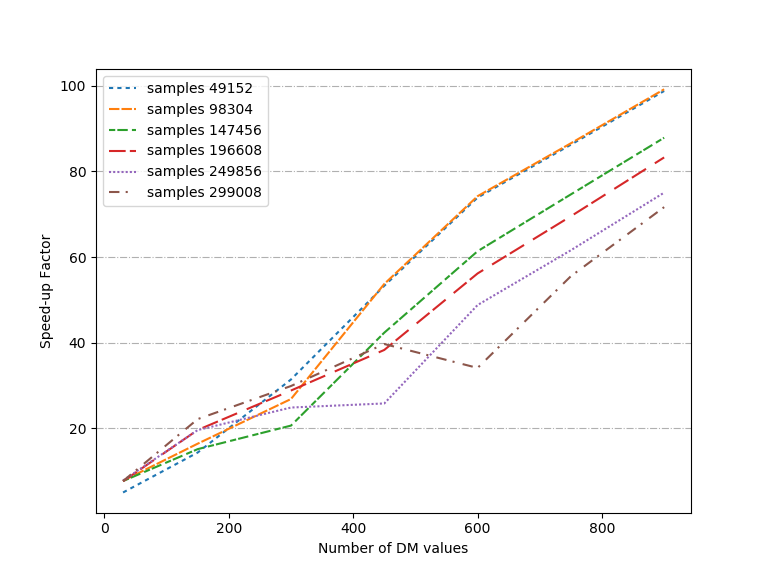} \\
\end{tabular}
\caption{When the number of channels is 256, 512, 1024, and 2048, the gpuprepsubband is compared with the acceleration effect of mpiprepsubband running 15
processes. It can be seen from the figure that as the number of tested DMs increases, the acceleration effect becomes better and better; when the number of
DMs is relatively large , The acceleration effect of short files is better than that of long files.}
\label{fg3:mpigpu}
\end{figure*}

The testing process was conducted on an AMD Ryzen 7 Threadripper 2700X with an NVIDIA GTX 1080 GPU, Table~\ref{tab:cpugpupara} shows the basic parameters of CPU and GPU. The operating system was Ubuntu 18.04 with CUDA Version 10.0, and the testing data was stored on the SSD. Figure~\ref{fg2:cpugpu} shows the speed-up effect of GPU relative to CPU single thread when the number of channels is 256, 512, 1024, 2048. As the number of tested DMs increases, the acceleration effect gets better and better, when the number of DMs is large enough, the acceleration effect of short files is better than that of long files. Figure~\ref{fg3:mpigpu} shows the speed-up effect of GPU relative to mpiprepsubband with the same parameters of CPU single thread testing, the general tendency of mpiprepsubband with GPU test was similar to the CPU single thread with GPU results.

The results of prepsubband are single-precision floating-point numbers recorded in dat files of binary format, within the allowable range of calculation accuracy, we checked the results by linux 'diff' command, and it was shown that the GPU results are the same as CPU.

In the early analysis and testing process of prepsubband, we found that Algorithm~\ref{alg2} takes up the largest proportion of calculation time, and the optimization of this part will not affect the effect of various options of prepsubband. We have also tested the 'mask','clip' and 'nobary' options, where the results of running on the CPU and GPU are consistent.

\begin{table}
\begin{center}
\caption{\textcolor{blue}{Basic parameters of CPU and GPU}}
\vspace{-5pt}
\begin{tabular}{l c c c c }
\hline
 \bf Processor type & \multicolumn{2}{c}{\bf CPU} & \multicolumn{2}{c}{\bf GPU} \\ \hline
\bf Version & \multicolumn{2}{c}{AMD Ryzen 7 2700X} & \multicolumn{2}{c}{NVIDIA GeForce GTX 1080} \\
\bf Core number & \multicolumn{2}{c}{8 cores, 16 threads} & \multicolumn{2}{c}{2560 cores} \\
\bf Clock rate & \multicolumn{2}{c}{3.7GHz} & \multicolumn{2}{c}{1.6/1.7GHz} \\
\bf Memory & \multicolumn{2}{c}{64GB} & \multicolumn{2}{c}{8GB} \\
\bf Compute ability & \multicolumn{2}{c}{-----} & \multicolumn{2}{c}{6.1} \\
\hline
\end{tabular}
\vspace{-10pt}
\label{tab:cpugpupara}
\end{center}
\end{table}

\subsection{Confirmation of 8 known pulsars with cross-matching in GSP $\&$ Parkes database}
\begin{figure*}
\begin{tabular}{cc}
\includegraphics[width=7.2cm,angle=0]{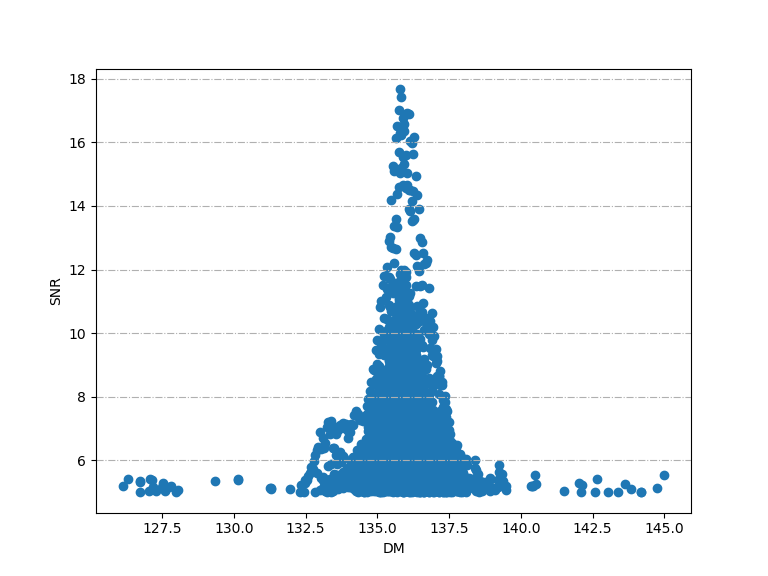} &
\includegraphics[width=7.2cm,angle=0]{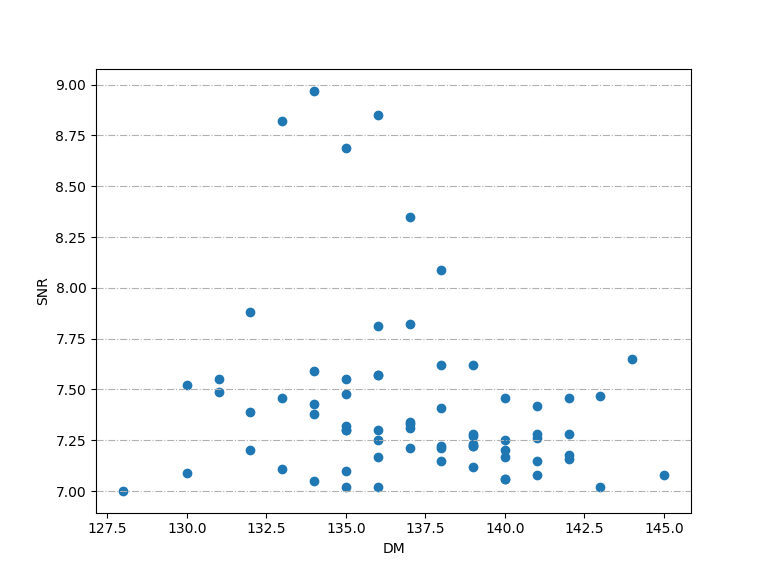} \\
J1823-0154 detected by FAST   &   J1823-0154 detected by Parkes \\
\includegraphics[width=7.2cm,angle=0]{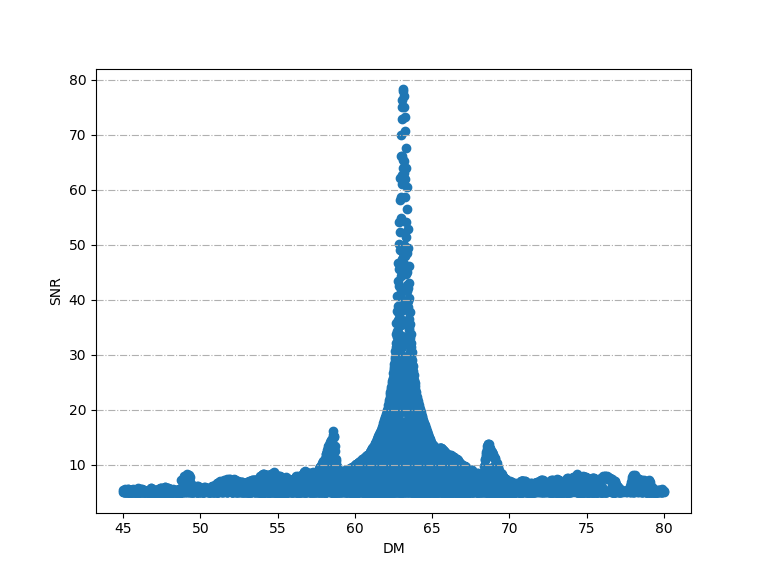} &
\includegraphics[width=7.2cm,angle=0]{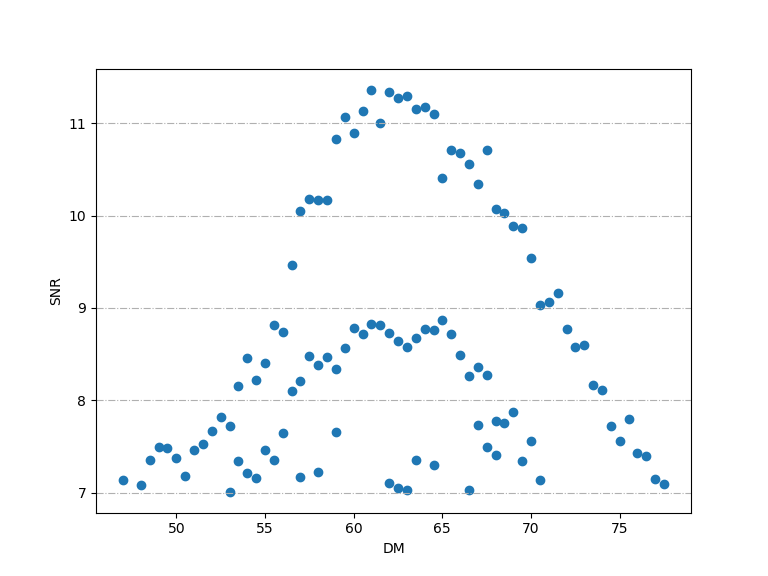} \\
J1824-0127 detected by FAST   &   J1824-0127 detected by Parkes \\
\includegraphics[width=7.2cm,angle=0]{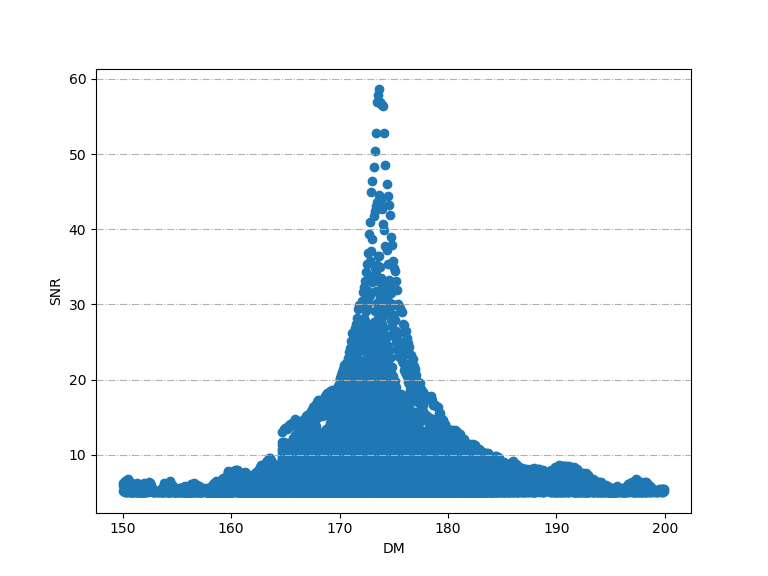} &
\includegraphics[width=7.2cm,angle=0]{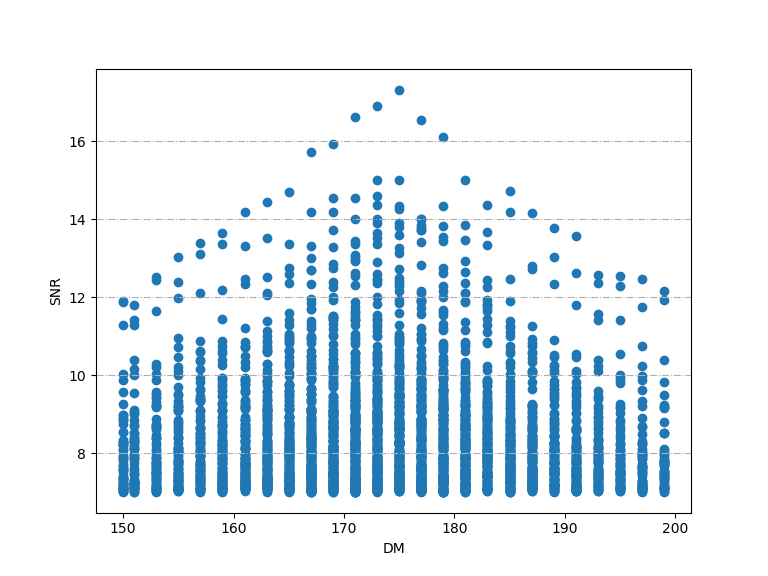} \\
J1852-0635 detected by FAST   &   J1852-0635 detected by Parkes \\
\includegraphics[width=7.2cm,angle=0]{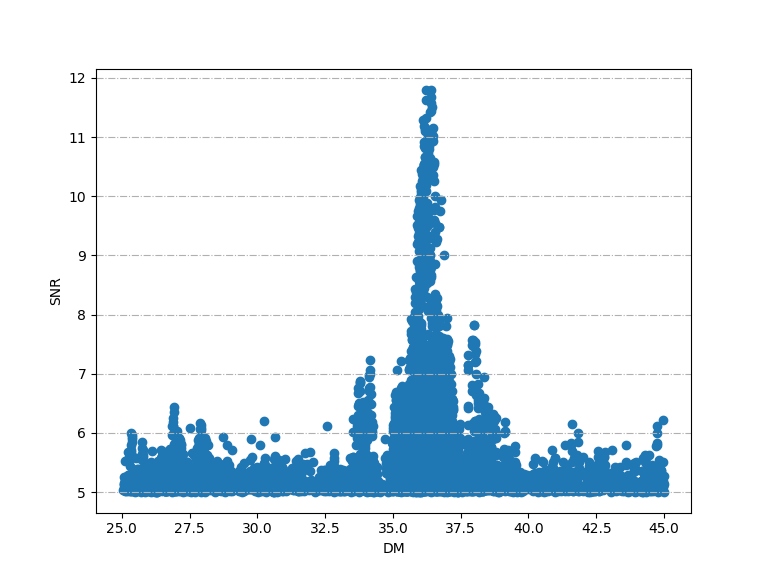} &
\includegraphics[width=7.2cm,angle=0]{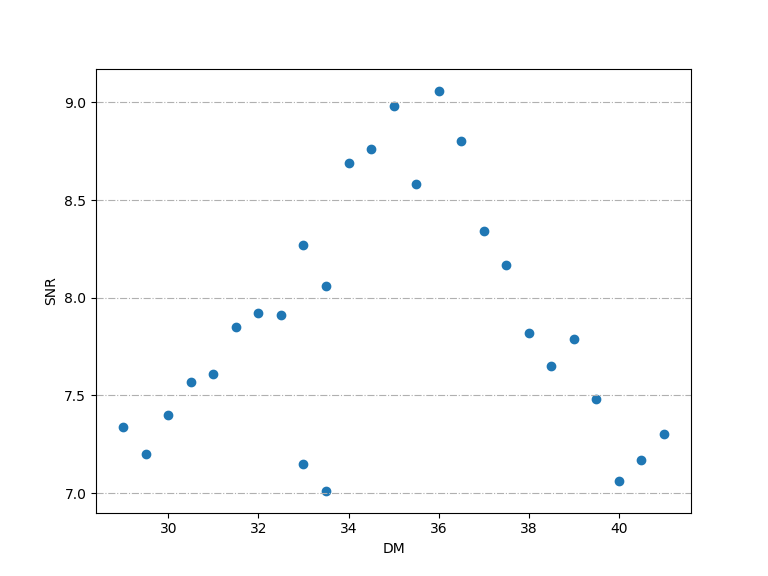} \\
J2005-0020 detected by FAST   &   J2005-0020 detected by Parkes \\
\end{tabular}
\caption{Comparisons of DM vs SNR plots of the 8 pulsars both detected by FAST and Parkes.}

\label{fg4:fastparkes}
\end{figure*}

We used the servers of the FAST early scientific data center to perform a single pulse search on the data of Band1 and Band2. After single pulse identification with RRATtrap.py in presto, 2244298 single pulse candidates have been classiled and recoded into database. Of the classified candidates,
1469415 have been classified as being RFI or noise, 100791 as excellent astrophysical candidates(with group rank 6), 674092 candidates with group
rank in 3,4,5. In these candidates, we compared the recordes of FAST UWB single pulse database with Parkes database and got 8 known pulsar both detected by FAST and Parkes~\citep{2020ApJS..249...14Z}, as shown in Table~\ref{tab:pulsars} and Fig~\ref{fg4:fastparkes}. For pulsar B1845-01, since the FAST drifting scan is far from the pulsar's position, the signal-to-noise ratio detected by FAST was weaker than that of Parkes.

\begin{table}
\begin{center}
\scriptsize
\caption{Comparison of 8 pulsars both detected by FAST and Parkes}
\vspace{-5pt}
\begin{tabular}{c c c c c c c c}
\hline
  \bf \text{No.} & \bf \text{Name} & \bf \text{P0} & \bf \text{DM} & \bf \text{Number of pulses} & \bf \text{Number of pulses} & \bf \text{Max SNR of pulses} & \bf \text{Max SNR of pulses} \\
  ~ & ~ & ~ & ~ & \bf \text{detected by FAST} & \bf \text{detected by Parkes} & \bf \text{detected by FAST} & \bf \text{detected by Parkes} \\ \hline
1 & B1822-09 & 0.769006 & 19.38 & 83714 & 76403 & 35.35 & 20.26 \\
2 & B1839+09 & 0.381319 & 49.16 & 101928 & 296 & 53.86 & 9.28 \\
3 & B1845-01 & 0.659432 & 159.53 & 13431 & 168190 & 12.42 & 22.17 \\
4 & B1846-06 & 1.451319 & 148.17 & 345565 & 11054 & 79.3 & 30.38 \\
5 & J1823-0154 & 0.759777 & 135.87 & 3421 & 70 & 17.66 & 8.97 \\
6 & J1824-0127 & 2.499468 & 58 & 14235 & 109 & 78.22 & 11.36 \\
7 & J1852-0635 & 0.524151 & 171 & 24782 & 3711 & 58.6 & 17.33 \\
8 & J2005-0020 & 2.279661 & 35.93 & 5077 & 27 & 11.8 & 9.06 \\
\hline
\end{tabular}
\vspace{-10pt}
\label{tab:pulsars}
\end{center}
\end{table}

\begin{figure*}

\begin{tabular}{cc}
\includegraphics[width=7.2cm,angle=0]{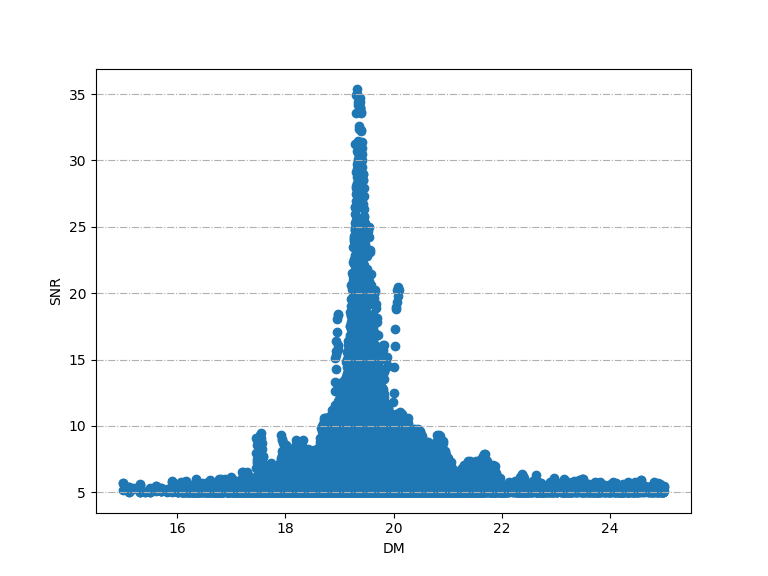} &
\includegraphics[width=7.2cm,angle=0]{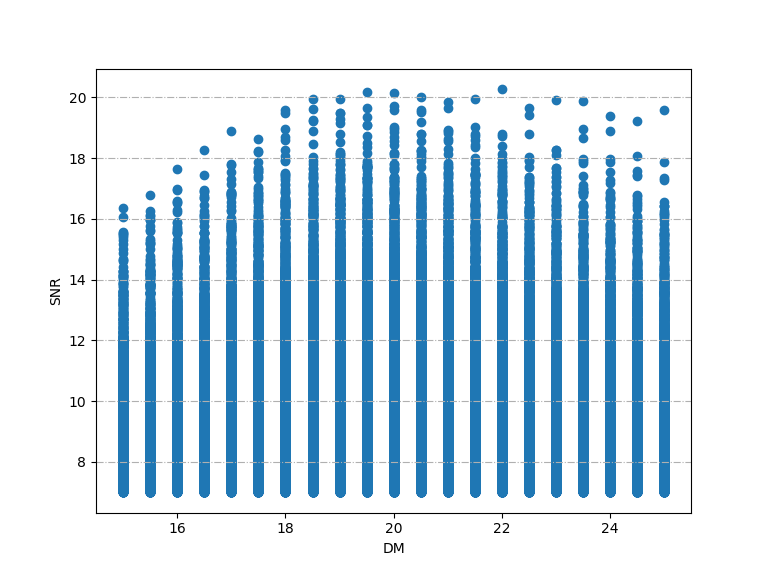} \\
B1822-09 detected by FAST   &   B1822-09 detected by Parkes \\
\includegraphics[width=7.2cm,angle=0]{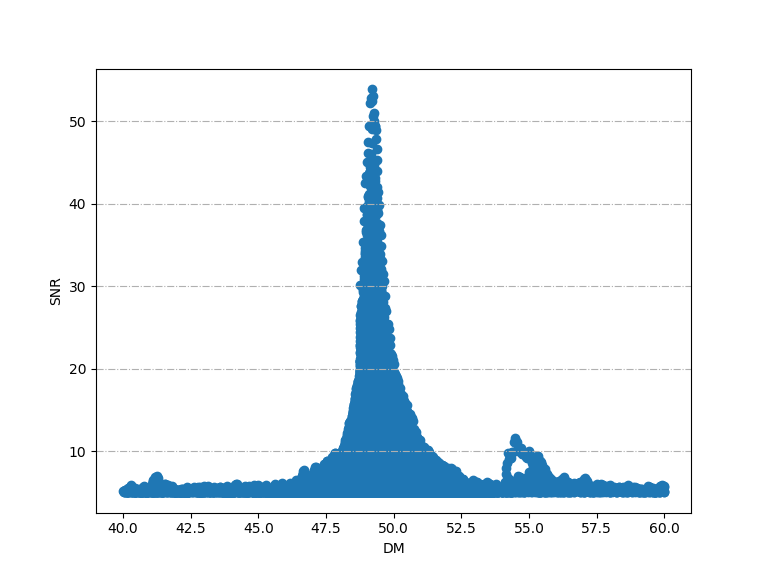} &
\includegraphics[width=7.2cm,angle=0]{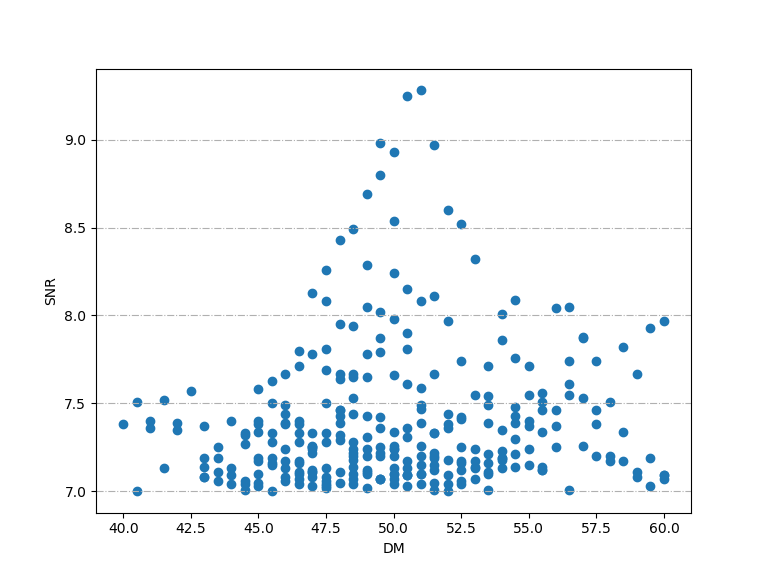} \\
B1839+09 detected by FAST   &   B1839+09 detected by Parkes \\
\includegraphics[width=7.2cm,angle=0]{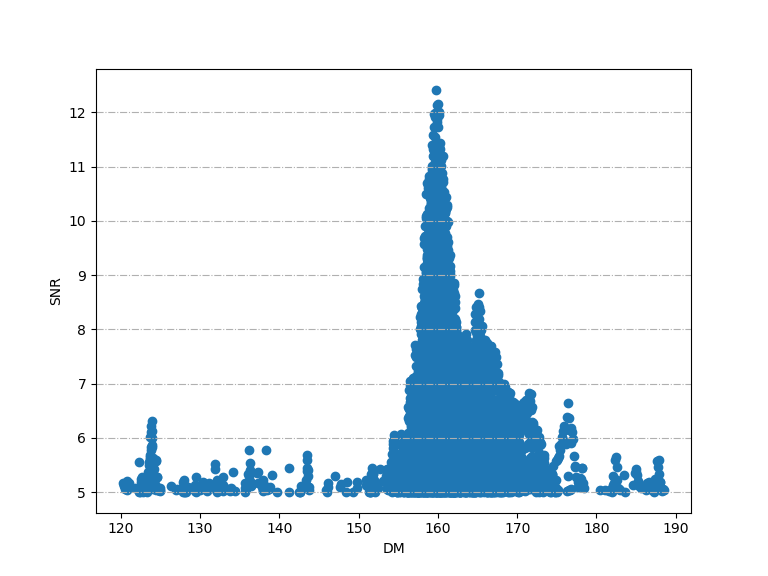} &
\includegraphics[width=7.2cm,angle=0]{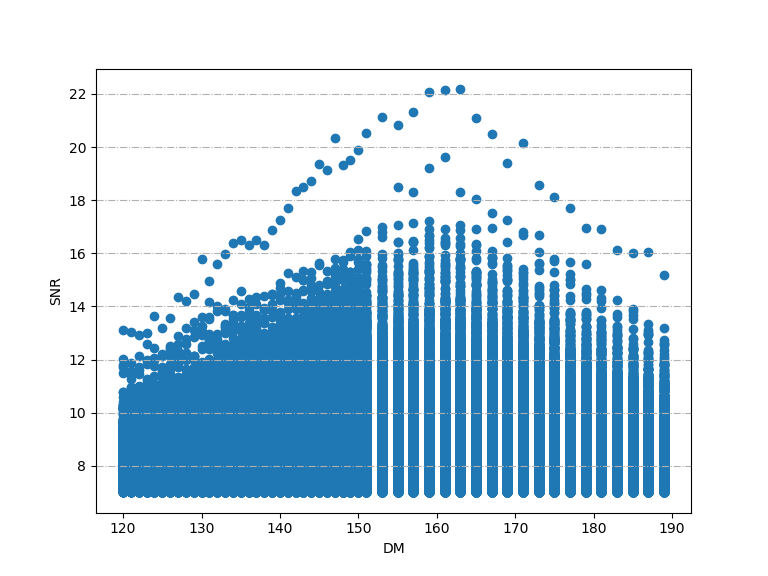} \\
B1845-01 detected by FAST   &   B1845-01 detected by Parkes \\
\includegraphics[width=7.2cm,angle=0]{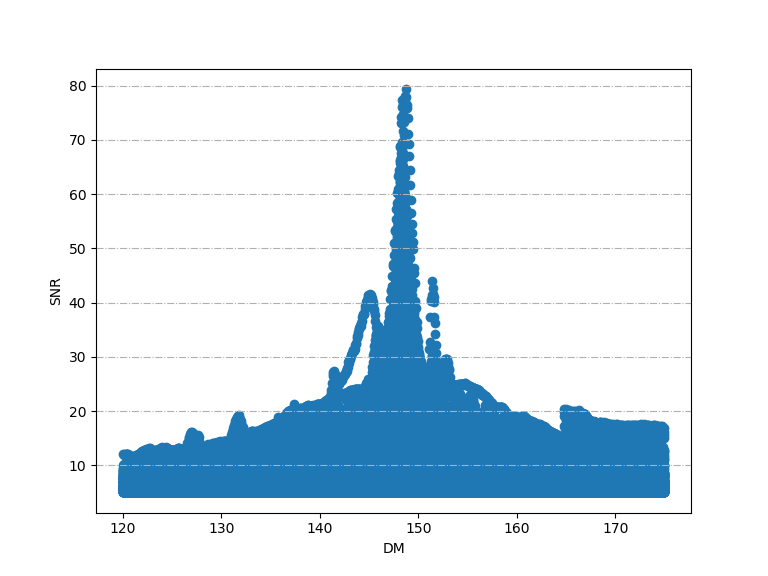} &
\includegraphics[width=7.2cm,angle=0]{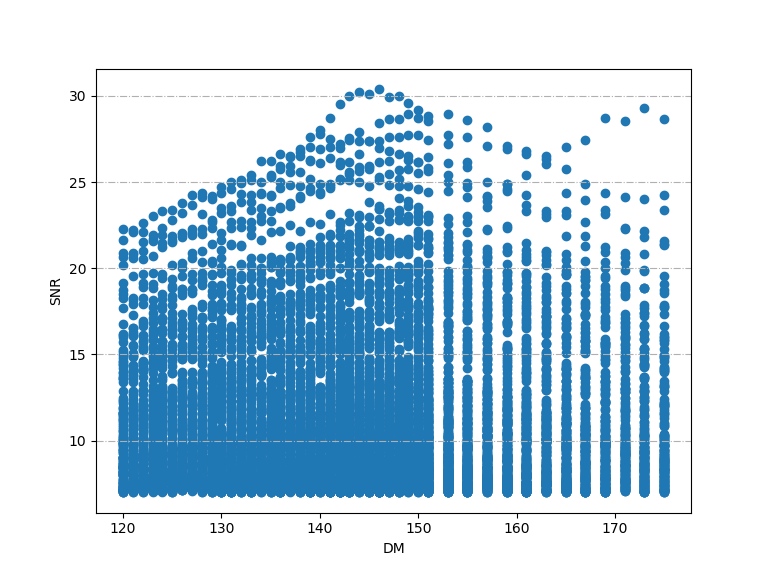} \\
B1846-06 detected by FAST   &   B1846-06 detected by Parkes \\
\end{tabular}
\end{figure*}

\begin{table}
\begin{center}
\scriptsize
\caption{13 new pulsars discovered by GSP}
\vspace{-5pt}
\begin{tabular}{c c c c c c c c c c}
\hline
  \bf \text{No.} & \bf \text{Name} & \bf \text{P0} & \bf \text{DM} & \bf \text{RA(2000)} & \bf \text{Dec(2000)} & \bf \text{GL} & \bf \text{GB} & \bf \text{D$_{NE2001}$} & \bf \text{D$_{YMW16}$} \\
  ~ & ~ & (s) & (cm$^{-3}$pc) & h:m:s & $\pm$d:m:s & ($^{\circ}$) & ($^{\circ}$) & (kpc) & (kpc)
  \\ \hline
1	&	J1931-0144	&	0.593661359782(12)	&	38.3(13)	&	19:31:32.025(9)	&	-01:44:22.5(4)	&	35.974	&	-9.711	&	1.7	&	1.4  \\
2	&	J1926-0652	&	1.608816302697(18)	&	85.3(7)	&	19:26:37.041(6)	&	-06:52:42.7(4)	&	30.751	&	-10.936	&	2.9	&	5.3  \\
3	&	J2323+1214	&	3.75949148736(8)	&	29.0(3)	&	23:23:21.619(7)	&	+12:14:12.70(16)	&	91.595	&	-45.209	&	1.7	&	\textgreater25  \\
4	&	J0402+4825	&	0.512194448728(3)	&	85.7(3)	&	04:02:40.633(9)	&	+48:25:57.51(7)	&	152.4276	&	-3.1772	&	2.3	&	1.8  \\
5	&	J2129+4119	&	1.68741829528(1)	&	32(1)	&	21:29:21.46(4)	&	+41:19:55(1)	&	87.1948	&	-7.1093	&	2.3	&	1.9  \\
6	&	J0529-0715	&	0.689223601359(8)	&	87.3(4)	&	05:29:08.973(2)	&	-07:15:26.43(10)	&	210.065	&	-21.582	&	\textgreater46	&	7.0  \\
7	&	J0021-0909	&	2.31413082912(16)	&	25.2(10)	&	00:21:51.47(3)	&	-09:09:58.7(11)	&	100.290	&	-70.726	&	1.3	&	\textgreater25  \\
8	&	J0803-0942	&	0.571256559156(14)	&	21.1(3)	&	08:03:26.848(13)	&	-09:42:50.81(14)	&	230.070	&	11.197	&	1.3	&	0.8  \\
9	&	J1502+4653	&	1.75250806373(3)	&	26.6(5)	&	15:02:19.83(1)	&	+46:53:27.4(1)	&	79.3056	&	57.6263	&	1.5	&	25.0  \\
10	&	J2112+4058	&	4.0607548114(5)	&	129(8)	&	21:12:51.76(6)	&	+40:58:04(1)	&	84.7385	&	-5.1570	&	5.4	&	5.2  \\
11	&	J2006+4058	&	0.499694912119(2)	&	259.5(2)	&	20:06:39.098(3)	&	+40:58:53.48(3)	&	77.1185	&	4.7313	&	\textgreater50	&	12.6  \\
12	&	J1951+4724	&	0.18192759882(1)	&	104.35(8)	&	19:51:07.45(3)	&	+47:24:35.1(2)	&	81.1828	&	10.3473	&	6.0	&	9.0  \\
13	&	J1822+2617	&	0.591417585722(1)	&	64.7(1)	&	18:22:44.819(2)	&	+26:17:26.83(4)	&	54.0268	&	17.5535	&	3.6	&	7.8  \\

\hline
\end{tabular}
\vspace{-10pt}
\label{tab:disc_pulsars}
\end{center}
\begin{tablenotes}
 \footnotesize
 \item The parameter are obtained from the timing solutions using the 64-m Parkes radio telescope ~\citep{2020MNRAS.495.3515C} and the 100-m Effelsberg radio telescope (Cruces 2021 MNRAS accepted, Wang Shen 2021 RAA accepted)
\end{tablenotes}
\end{table}

\begin{figure*}\centering
\includegraphics[width=12cm,angle=0]{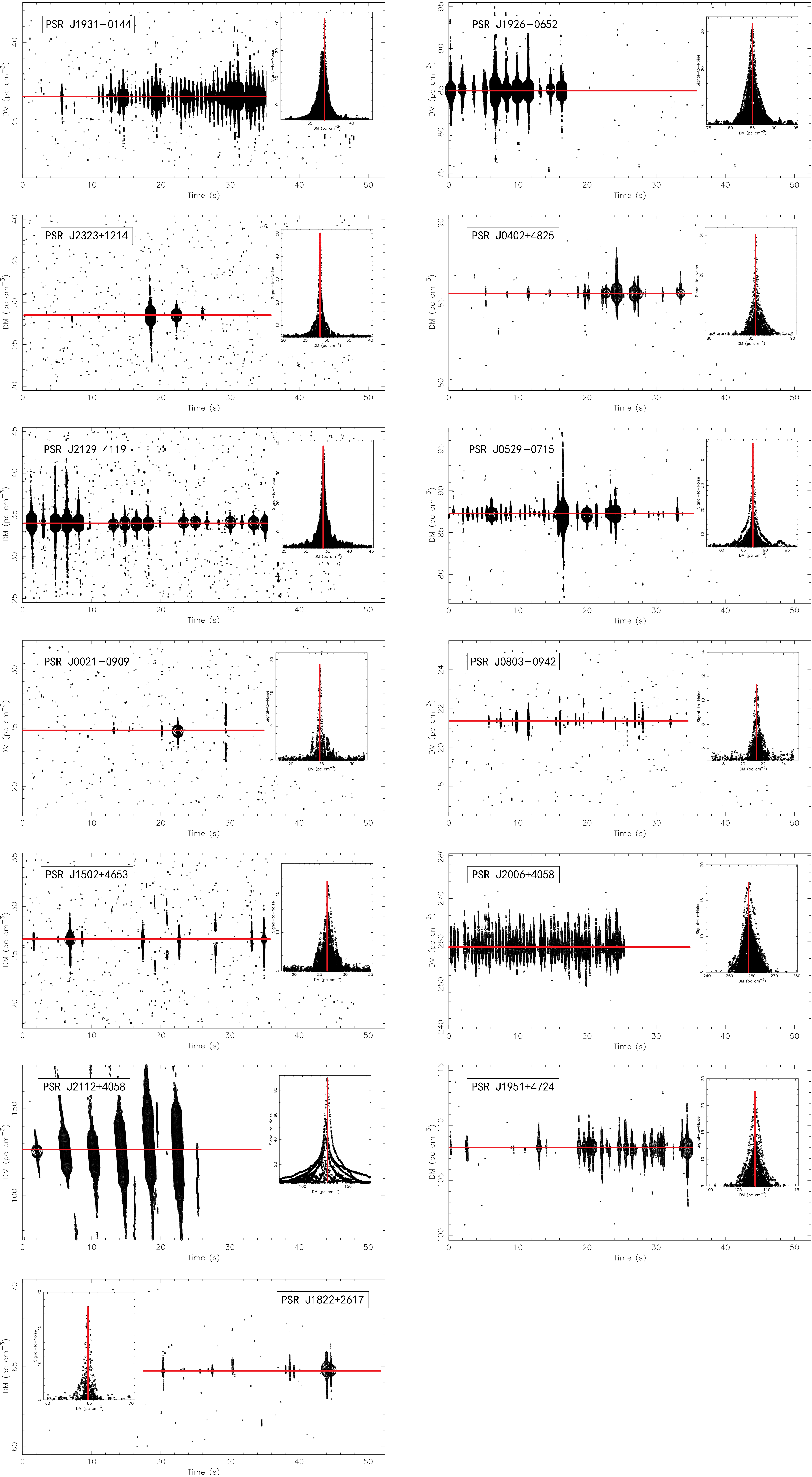}
\caption{Singlepulse search of 13 FAST pulsars discovered by GSP. The subpolts are scatter plots of the DM versus time for each single-pulse event whose is proportional to its S/N, the orizontal red line shows the DM value corresponding to the maximum S/N. Subplot are scatter plots of S/N versus DM, the vertical red line shows the DM value corresponding to the maximum S/N. }
\label{fg5:spresult}
\end{figure*}

\section{Discovery of 13 pulsars from CRAFTS survey}

Using GSP pipeline, we discovered 13 new pulsars\footnote{https://crafts.bao.ac.cn/pulsar/} as shown in Table ~\ref{tab:disc_pulsars}, and used FFA ~\citep{1969IEEEP..57..724S} for periodical search. For the single pulses adjacent to the DM, the maximum range of the FFA search is 1.5 times the time interval of the two single pulses adjacent to the DM; the follow-up timing observation has been completed by Prakes and Effelsberg Radio Telescope, these pulsars have been reported upon ~\citep{2020MNRAS.495.3515C,2019SCPMA..6259508Q,2019ApJ...877...55Z} and (Cruces 2021 MNRAS accepted, Wang Shen~\citep{2021RAA....21...91Y})), the discovery plots of GSP for each of 13 pulsars are shown in Figure~\ref{fg5:spresult}.

\clearpage
\section{Conclusions}
\label{sect:conclusions}

We here described a new, more efficient single pulse search pipeline, namely GSP,  to search for pulsars, RRATs, and FRBs. The pipeline uses
GPU to accelerate dedispersion, and use CPU parallel threading to implement single pulse search, single pulse identification, and folding processes. Our main results are

\begin{itemize}
\item GSP integrates data preparation, dedispersion, single pulse search, candidate ranking, and candidate archiving.
\item GSP implement the presto dedispersion package 'prepsubband' with CUDA, which was shown to be  $\sim$120 times faster than the original 'prepsubband' and $\sim$60 times faster than the MPI version for processing CRAFTS data.
\item We detected all known pulsars in the overlapping sky of the FAST-UWB observation$\&$ Parkes Multibeam surveys.
\item GSP discovered 13 new FAST pulsars in the pilot CRAFTS survey with the FAST-UWB.
\end{itemize}

\clearpage
\normalem
\begin{acknowledgements}
This work is supported by National Natural Science Foundation of China (NSFC) Programs No.  11988101, No.  11725313, No.  11690024, No.12041303 No.  U1731238, No.  U2031117, No. U1831131, No. U1831207 and supported by Science and Technology Foundation of Guizhou Province Programs No. LKS[2010]38; PW acknowledges support by the Youth Innovation Promotion Association CAS (id. 2021055) and cultivation project for FAST scientific payoff and research achievement of CAMS-CAS. We thank the anonymous referee for useful comments on the manuscript. This work made use of data from the Five-hundred-meter Aperture Spherical radio Telescope (FAST), FAST is a Chinese national mega-science
facility, built and operated by the National Astronomical Observatories, Chinese Academy of Sciences.

\end{acknowledgements}

\bibliographystyle{raa}
\bibliography{bibtex}

\end{document}